# Market-Neutral Strategies in Mid-Cap Portfolio Management: A Data-Driven Approach to Long-Short Equity

Saumya Kothari, Harsh Shah, Utkarsh Prajapati, Shrinjay Kaushik
*s.b.kothari@columbia.edu, harsh.s@columbia.edu, up2147@columbia.edu, shrinjay.kaushik@columbia.edu*

*Abstract*—Mid-cap companies, generally valued between *$2 billion and $10 billion*, provide investors with a well-rounded opportunity between the fluctuation of small-cap stocks and the stability of large-cap stocks. This research builds upon the long-short equity approach (e.g., Michaud, 2018; Dimitriu, Alexander, 2002) customized for mid-cap equities, providing steady risk-adjusted returns yielding a significant Sharpe ratio of 2.132 in test data. Using data from 2013 to 2023, obtained from WRDS and following point-in-time (PIT) compliance, the approach guarantees clarity and reproducibility. Elements of essential financial indicators, such as profitability, valuation, and liquidity, were designed to improve portfolio optimization. Testing historical data across various markets conditions illustrates the stability and resilience of the tactic. This study highlights mid-cap stocks as an attractive investment route, overlooked by most analysts, which combine transparency with superior performance in managing portfolios.

*Index Terms*—market neutral strategy, mid-cap stocks, portfolio management, long/short equity

## I. INTRODUCTION

Mid-cap equities provide a distinctive mix of reliability and growth opportunity. These firms often face critical phases in their development paths, weighing the unpredictability of smaller businesses against the stability of larger, more established companies. For investors, mid-cap stocks offer a chance to benefit from the expansion of up-and-coming leaders while effectively controlling risk (e.g., Ge, 2018; Nayan, 2022). However, these stocks are often *underrepresented* in portfolios, leading to inefficiencies and hidden opportunities for data-driven strategies.

This paper aims to create a market-neutral investment approach intended to take advantage of inefficiencies in mid-cap equities. Using a long-short equity strategy, the approach reduces exposure to general market movements while focusing on alpha generated from individual stock attributes (Sorensen, Ma et al. 2006; Francis, LaFond et al. 2004). By using historical data from 2013 to 2023 and adhering to point-in-time (PIT) compliance to prevent data leakage, our analysis aligns closely with real-time investor decision-making.

The paper has been structured in the following sections: Section II discusses the Relevant Review of Literature from academicians and industrialists who have begun to explore mid-cap companies and market-neutral strategies. This section also addresses the gap which our research aims to fill. Section III speaks about the data, and the process we have gone about collecting it, whereas Section IV dives deep into the methodology we have followed to analyze the data and run our algorithm of creating a dollar neutral portfolio with robust training, testing & validation. Section V and VI talk about the results and limitations of the research conducted. Finally, Section VII speaks about future scope of how further research can be built of the premise which this research study has laid, with our concluding thoughts in Section VIII.

## II. RELATED WORK AND MAIN CONTRIBUTIONS

### A. *Market Neutral Strategies*

Market-neutral strategies are widely known to mitigate risk by not completely relying on the overall movement or "wave" of the market. This strategy is particularly noteworthy to generate positive alpha by exploiting various market inefficiencies which occur in stock price and other features as outlined by Nicholas (2000). It was described that equal and opposite direction, long and short positions balanced out the market exposure despite its fluctuations. AllianceBernstein (2012) further put weight to the market-neutral strategies emphasizing the reliance on manager expertise coupled with diversification of stocks - a method which has significantly gained traction during periods of high market stress.

In more recent times, Johannessen and Nordestedt (2023) explored the aspect of stock-picking in a systematic way rather than market exposure, wherein uncorrelated returns were generated. For this reason, this strategy has gained lots of attention in the recent years to maximize the mean-variance optimization.

### B. *Mid-Cap Investing*

In addition to market-neutral strategies, another sector which are exponentially growing in importance are the medium capitalized companies or "mid-cap" stocks. This segment has been widely overlooked in academia due to the *salience bias* - which refers to focusing only on the extreme values i.e. research on small-cap and large-cap stocks, and not focusing on the middle i.e. mid-cap stocks. However, these 'hidden gems', as proclaimed by Charlie Munger, have been increasingly gaining importance and appeal.

Lynch (2018) discovered that mid-cap firms are generating significantly higher 3-year average returns relative to small and large-cap stocks from 1982-2017 - this was a hallmark achievement bringing to light the amount of alpha that can be captured from investing into these. This study was highlighted in many works due to its challenge on the authority of the small-cap stocks and brought about the phenomenon of 'mid-cap anomaly'.

Kayne Anderson Rudnick (2023) further explored the case for mid-cap investing, highlighting the segment's large opportunity set and compelling return patterns compared to large-cap and small-cap stocks. Their research suggested that mid-caps retain qualities of both large-cap and small-cap stocks while offering unique advantages.

American Century Investments (2024) provided additional evidence supporting the out-performance of mid-cap stocks over extended periods. Their analysis demonstrated that mid-cap stocks outperformed large-caps in 55% of rolling five-year periods since 1983 and small-caps in 89% of such periods.

### C. Addressing gaps

Our study aims to bridge this gap by developing a robust long-short equity strategy tailored for mid-cap stocks. We build upon the existing literature in several key ways:

- The first contribution is the novel focus on midcap stocks, with relation to the market neutral strategy. There have been so many greats - Howard Marks, Charlie Munger - that speak about these medium capitalization-valued stocks being strong rooted in their growth potential, but rarely any academic literature about using exploring the value with these stocks.
- The second contribution is the optimization using long/short equity strategy. This dollar-zero strategy is crucial during extremely volatile times in the market, increasingly frequent with the passage of time. As times in recent years have become increasingly volatile (Zhang, Wei, 2006) with the occurenace of wars, increase of military power, changing governemnts - it is vital to underscore this method of trading to make consistent returns, no matter where the total market goes. In this paper, we dive deep into this market-neutral trading strategy.
- The third contribution is the paper develops a robust, data-driven market-neutral strategy tailored for mid-cap stocks. This approach combines:
  – Advanced feature engineering using profitability, valuation, and liquidity metrics
  – Strict point-in-time (PIT) compliance to prevent look-ahead bias
  – A dollar-neutral portfolio optimization framework
  This comprehensive methodology offers a new perspective on exploiting inefficiencies in the mid-cap market segment.

### III. DATA AND METHODS

#### A. Data Overview & Extraction

The information for this research was obtained from WRDS, namely, from the CRSP and Compustat databases, which encompass U.S. equity market information. The study takes into consideration, the period from January 2013 to December 2023, gathering monthly data on the first trading day of the following month. This method of financial data extraction guarantees that strictly just historically accessible data is employed, following a rigorous point-in-time (PIT) approach and preventing look-ahead bias.

Data preparation included multiple crucial stages. Corporate actions of stock split and declared dividends were modified to ensure uniformity among stock prices. Missing data points were 'filled in' solely using information from the past, without adding any future values. For instance, if earnings data for a specific month was unavailable or missing, the latest, previous valid earnings data was utilized until new information was received. This "forward fill only" guideline ensured data integrity without causing predictive data leaks.

Unique identifiers such as *'permno'* and *'gvkey'* were referenced to combine data from *'CRSP'* and *'Compustat'*. The merging procedure was attentively monitored to guarantee consistency among datasets and to prevent the integration of mismatched or duplicate entries. The final dataset underwent additional screening for irregularities, excluding observations with extreme or improbable values that could potentially skew the analysis for this study.

#### B. Data Integrity & Validation

To prevent any leakage of future information, the validation framework was carefully implemented to maintain temporal consistency between the training(2013-2021) and testing(2023) datasets. The data was divided into training, validation (2022), and testing phases. For the purpose of training, only historic, time-relevant data was utilized. Validation and testing datasets strictly conformed with the point-in-time implementation, ensuring that no financial information was leaked from upcoming, future years.

For example, when combining datasets, fields like earnings or book value were populated using the latest valid entries that were accessible at that moment. No anticipatory modifications or interpolations were carried out. Moreover, the use of a rolling window back-testing framework guaranteed that the predictions for each period were based exclusively on data accessible up to that moment.

By following these principles, the research guaranteed that the outcomes are both strong and repeatable, remaining consistent with actual investment situations.

#### C. Data Engineering

From the CRSP dataset we collected the stock prices, outstanding shares and the corresponding date data, which was then used for calculating the market capitalization over time for the stocks. These market capitalization values were filtered between the value of $2 Billion to $10 Billion to fit the definition of the Mid-Cap stocks. Having collected the Mid-Cap stocks over time from the above data, we filtered every subsequent data we collected for only the Mid-Cap stocks.

From the Compustat dataset, we collected the factors that collectively provided insights into a company's profitability (e.g., Net Income), financial health and stability (e.g., Debt, Equity, Cash, and Liabilities), operational efficiency (e.g., EBITDA, Operating Income), and liquidity (e.g., Current Assets, Liabilities). These are commonly used to get a comprehensive view of a company's profitability, gauge earnings potential, financial stability, to gauge risk and leverage, and operational efficiency, and to evaluate how well the company manages costs and generates core income. Liquidity metrics tell us about a firm's ability to meet obligations in the short-term, while revenue growth indicates a potential for future expansion. Together, these factors can help us identify financially strong, efficient, and growth-oriented companies, forming the backbone of a sound investment strategy focused on achieving good levels of risk-adjusted returns. A brief explanation of each of the factors reigned in from the dataset has been given below: ni: Net Income epspx: Earnings Per Share (Basic, Excluding Extraordinary Items) ceq: Common Equity (Stockholders' Equity) revt: Total Revenue dltt: Long-Term Debt dlc: Current Portion of Long-Term Debt (Short-Term Debt) che: Cash and Cash Equivalents ebitda: Earnings Before Interest, Taxes, Depreciation, and Amortization act: Current Assets lct: Current Liabilities oiadp: Operating Income After Depreciation xint: Interest Expense at: Total Assets

Using the above factors we were able to get a comprehensive list of financial ratios which would make the backbone of the data that forms our investment strategy in this study. The following ratios were calculated using the above factors: Further, From the CapitalIQ dataset, we collected the News

TABLE I: Financial Ratios and Their Calculations

| Financial Ratio | Calculation |
|---|---|
| P/E Ratio | Price-to-earnings ratio, calculated as `prc / epspx` |
| P/B Ratio | Price-to-book ratio, calculated as `prc / book-value-per-share` |
| EV/EBITDA | Enterprise value / EBITDA |
| Gross Margin | Gross profit as a percentage of revenue `gp / revt` |
| Operating Margin | Operating income as a percentage of revenue `oiadp / revt` |
| Net Margin | Net income as a percentage of revenue `ni / revt` |
| Current Ratio | Ratio of current assets to current liabilities `act / lct` |
| Debt-to-Equity Ratio | Ratio of total debt to shareholders' equity `lt / ceq` |
| Interest Coverage Ratio | Ratio of EBITDA to interest expense, a measure of solvency |

Data for different companies over time. This News Data was in the form of various news headlines relating to different companies over time.

We utilized Python's natural language "nltk" toolkit for Sentiment Analysis on these news headlines. Specifically, from the nltk library, the "SentimentIntensityAnalyzer" was utilized to perform sentiment analysis on the headlines. It computed a compound sentiment score for each headline, which measured the overall sentiment (either positive, negative, or neutral) on a scale from -1 to 1, with -1 denoting extremely negative sentiment, 1 extremely positive and 0 a neutral sentiment. This value was stored in the 'avg_sentiment' data field in the final dataset.

A detailed explanation of all the fields in the final dataset obtained is given in a table at the end of the report in the appendix section (section IX).

Finally, for merging data from the CRSP and Compustat dataset, we used the CRSP-Compustat link table to map "permno"s to "gvkey"s. Since CRSP uses the former as its unique identifier and CompuStat the latter. Further merging of the sentiment data from the CapitalIQ dataset was done on the gvkey(s) unique identifier as well.

IV. METHODOLOGY

Our investment strategy employs a data-driven, market-neutral approach, focusing on mid-cap stocks—defined as companies with market capitalization ranging between $2 billion and $10 billion. This range represents a segment of the market that combines growth potential with relative stability, making it ideal for systematic strategies aimed at exploiting market inefficiencies. The methodology comprises several key components that ensure the strategy remains robust, transparent, and adaptable to changing market dynamics.

A. Outlier Treatment

Outliers can lead to a lot of noise and distort statistical analyses and machine learning models. To arrive at precise results that figure out the true patterns and trends from the data, we had to address the outliers in the dataset. For this, we employed the following techniques:

- **Data Transformation**: Certain features such as the price-to-earnings ratios and the price-to-book ratio were leading to a lot of outliers. We utilized Reciprocal Transformation i.e. utilizing the inverses of these ratios in place - E/P and E/B ratios - which had more in-range values and thus contributed less noise to the dataset and also improved the statistical features and impact of those ratios.
- **Winsorization**: We applied the Z-score method from Python's Scipy library to all the numeric features in our dataset. This centered each feature around a zero mean with a standard deviation of 1. Outliers were then identified by looking at the values of features that had a Z-score of either less than -3 or more than 3, since these were the values that were 3 standard deviations away from the feature mean. These outliers were then "clipped" or "capped" so that their Z-score became -3 or 3, whichever value was closer. Interesting thing to note here was before clipping we employed imputing the outliers by the median value, but that did not get rid of the distortion in the data as effectively as clipping did.

By reducing the outliers in this way, we reduced the undue influence of extreme values and enhanced the dataset's representativeness.

## B. Multicollinearity Treatment

Multicollinearity, i.e. high correlations amongst the predictor variables in a dataset, can lead to unstable estimates and substantially reduce the explanatory power of the model, and thus needs to be addressed to enhance robustness and reliability of the optimization model. To address multicollinearity, we applied the following methods:

- **Variance Inflation Factor (VIF)**: Features with VIF values greater than 10 were eliminated to ensure minimal redundancy in the predictors. A high VIF indicates that a particular variable is strongly correlated with others, which can distort the optimization results.
- **Correlation Matrix Analysis**: Pairwise correlations were analyzed, and variables with correlation coefficients exceeding 0.8 were flagged for redundancy. Highly correlated features were visually represented in a correlation matrix, as shown in Fig. 1.
- **Representative Variable Selection**: For each group of highly correlated variables, a single representative feature was retained based on its relevance to the strategy, interpretability, and contribution to the overall performance of the model. This step ensured that only the most meaningful features were included while preserving the predictive power of the data.

By systematically reducing multicollinearity, we improved the stability of the model and mitigated the risks of over-fitting, particularly when dealing with smaller subsets of mid-cap stocks.

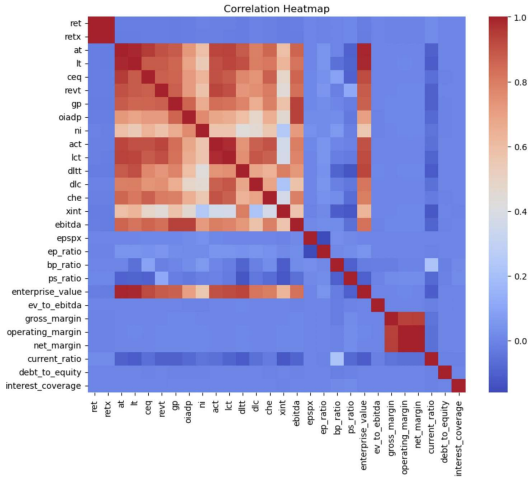

Fig. 1: A correlation matrix for the features identified from the data

## C. Financial Factors

Following multicollinearity treatment, the remaining features were categorized into three primary groups of financial metrics, each designed to capture different dimensions of company performance:

- Profitability Metrics: These are the variables that assess the ability of a company for generating earnings relative to its revenue or other benchmarks. Examples include:
  - *Net Income*: The bottom-line profit after all the expenses, taxes, and interest have been adjusted for.
  - *Gross Margin*: Is the measure of how efficiently a company produces goods; its calculated as the gross profit over the total revenue.
  - *Earnings per Share (EPS)*: Is a key indicator of profitability; it reflects the portion of the net income that is attributed to each outstanding share.
- Valuation Metrics: These are the metrics that provide information on the company's stock price value relative to its fundamentals. Key valuation measures include:
  - *Earnings-to-Price Ratio (E/P)*: This reflects the earnings that is generated per dollar of the stock price; its useful for identifying undervalued companies.
  - *Book-to-Price Ratio (B/P)*: This is a measure of the price of the stock compared to its book value; its often used in deep-value strategies.
  - *Enterprise Value to EBITDA*: Compares enterprise value to earnings before interest, taxes, depreciation, and amortization; it provides a cleaner measure of the company's performance.
- Liquidity and Solvency Measures: These variables are useful in assessing a company's financial health and also its ability to be able to meet obligations. Examples include:
  - *Current Ratio*: This ratio is calculated as the current assets to the current liabilities, which gives a measure of the short-term liquidity.
  - *Debt-to-Equity Ratio*: The proportion of debt used relative to equity; its useful in measuring financial leverage.
  - *Interest Coverage Ratio*: Reflects the ability to cover interest expenses originating from operating profits; its thus a critical measure of solvency.

These financial factors form the backbone of identifying profitable opportunities within the mid-cap avenue, ensuring the strategy maintains alignment with core fundamental analysis principles.

## D. Portfolio Construction

The core facet of the strategy is constructing a dollar-neutral portfolio that strikes a balance between long and short positions thus minimizing exposure to systematic market risks. The optimization framework's objective is to maximize the risk-adjusted return of the portfolio and is thus mathematically expressed as:

$$\max_{\mathbf{w}} \mathbf{w}'\boldsymbol{\mu} - A\mathbf{w}'\boldsymbol{\Sigma}\mathbf{w} \quad (1)$$

subject to the constraint:

$$\mathbf{w}'\mathbf{1} = 0 \quad (2)$$

where:
- $\boldsymbol{\mu}$: vector of expected returns for each asset,
- $\boldsymbol{\Sigma}$: variance-covariance matrix of asset returns,
- $A$ is the risk-aversion parameter (set to 2 for this strategy),

- **w** is the weight vector representing portfolio allocations.

The dollar-neutral constraint ensures that the sum of long positions equals the sum of short positions, isolating alpha generation from market-wide movements. By simultaneously going long on undervalued assets and short on overvalued ones, the strategy maintains neutrality to broader market trends while capitalizing on stock-specific opportunities.

### E. Risk Management

Risk management is an integral part of the strategy, ensuring that portfolio exposures remain controlled and aligned with the market-neutral objective. The following components are central to the risk management framework:

- **Dollar Neutrality**: As illustrated in Fig. **2**, the constraint ensures:

$$\sum_{i \in \text{Long}} w_i^L \cdot P_i = \sum_{j \in \text{Short}} w_j^S \cdot P_j \qquad (3)$$

where, $w_i^L$ and $w_j^S$ represent the weights of the long and short positions, respectively, and P is the price of the asset.

- **Alpha Independence**: By neutralizing systematic market exposure, the portfolio generates alpha purely from factor-driven signals, independent of market trends.

- **Controlled Risk Factors**: Exposure to specific risk factors, such as liquidity, sector concentration, and volatility, is continuously monitored to prevent unintended imbalances. This framework allows for alpha generation independent of broader market movements while maintaining controlled exposure to specific risk factors.

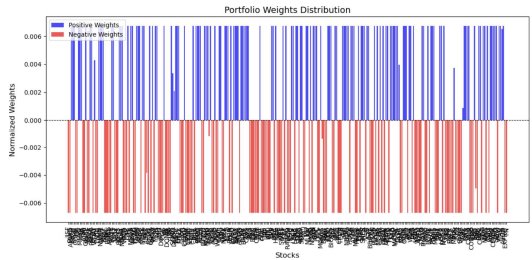

Fig. 2: Optimized dollar-neutral portfolio weights (blue = long, red = short, height = weight values))

## V. RESULTS

The objective of this paper was to evaluate the dollar neutral portfolio approach to mid-cap stocks present in the universe of U.S. equities. We conducted an avid process of filtering the data by removing outliers through winsorization and creating financial ratio columns from the data we had present from WRDS. We perform a rigorous evaluation process where we have an initial training dataset with data from 2013-2021, which we validate on data from 2022. After this, we re-train our model on data from 2013-2022, and finally test it out on 2023 data. This rigorous training and re-training occurred to make the model even more efficient.

We made sure our data was certainly Point-in-Time compliant while running this algorithm. We made sure that no backward filling was allowed, and only results were computed on a forward fill and then re-adjusted according to compute the forward fill on the re-trained data. To prevent the look ahead bias - where future information inadvertently affects past results to make them seem more appealing - we only used historical data that was available up to each specific point in time, ensuring that no data or signals from future periods influenced past observations or decisions. For example, when calculating financial metrics such as returns, beta values, and other indicators, we employed a rolling-window approach while maintaining chronological consistency. Our back-testing methodology aimed to validate the strategy's robustness while avoiding look-ahead bias and overfitting. We implemented a three-phase testing approach:

A. *Initial Model Training (Period: 2012–2021)*
   • Used to establish baseline model parameters and factor weights.
   • In-sample Sharpe ratio: **2.59681**.
   • Optimized portfolio weights using the dollar-neutral constraint.

B. *Validation Phase (Period: 2022)*
   • Used to test model stability and parameter sensitivity.
   • Out-of-sample Sharpe ratio: **2.3914**.
   • Allowed for model refinement and hyper parameter tuning.
   • Helped identify potential over-fitting issues.
   • Model retraining incorporated validation feedback.

C. *Out-of-Sample Testing (Period: 2023)*
   • Applied retrained model to completely unseen data.
   • Out-of-sample Sharpe ratio: **2.13211**.
   • Portfolio maintained dollar neutrality throughout testing.
   • Demonstrated strategy's real-world applicability, with high returns compared to benchmarks like the S&P 500, as seen in Fig. 3.

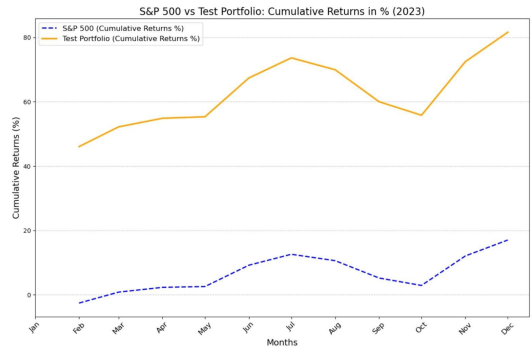

Fig. 3: Comparison of out of sample test portfolio returns to S&P500 returns for 2023)

An important observation to note is the weights given to every stock are primarily occurring in the range of (-0.006, +0.006). This seems interesting from the point of view of the massive diversification of funds which has occurred to reduce volatility - an aspect very in line with creating a robust market-neutral strategy. On the other hand, to require further investigation if the weights may be very small put on the stocks, thus hindering further alpha to be captured if weights are optimized further.

The results seem promising but steadily decline as we approach the testing data, which is somewhat expected. It may be a valid to make this assumption that more the transaction and trading costs increase the algorithm will perform with lower Sharpe ratio than calculated. Furthermore, testing this strategy in more volatile times historically may further reveal insights into the strength, however, in the current time period observed, the results seem promising. The main reason for this success is the rolling window approach which allowed continuous model adaptation while avoiding data leakage. The strong performance metrics, particularly the out-of-sample Sharpe ratio, show the strategy is effective in capturing mid- cap market inefficiencies.

## VI. Discussion

The strategy demonstrates high returns via systematic factor exploitation, adjusting for risk with a long-short approach to ensure alpha remains independent of broader market movements. However, there are limitations when implementing this strategy for real-world trading.

### A. Technical Limitations

*Parameter Sensitivity:*

- The strategy shows significant sensitivity to the risk-aversion parameter ($A = 2$) in the optimization framework. We have used this risk aversion parameter as one of our assumptions, and by changing the value can greatly change the optimization.
- Portfolio weights are highly dependent on accurate estimation of factor importance, making the model vulnerable to estimation errors.

*Market Condition Dependencies:*

- The back-testing period may not fully capture:
  - Extreme market scenarios,
  - Various economic cycles,
  - Different interest rate environments,
  - Regime changes that could affect mid-cap performance.

### B. Implementation Constraints

*Trading Frictions:*

- The strategy faces real-world challenges, including:
  - Transaction costs from frequent re-balancing,
  - Market impact when trading less liquid mid-cap stocks,
  - Potential slippage during execution.

*Portfolio Construction:*

- The dollar-neutral constraint requires:
  - Maintaining equal long and short positions,
  - Continuous re-balancing to maintain neutrality,
  - Available short inventory for selected stocks.

- Most of the above limitations of the algorithm we have used have been mentioned above. However, the 'Implementation Constraints' can significantly hurt the alpha of the strategy. Keeping these factored into the model can make the accuracy of replicating real-world scenarios higher.

## VII. Future Work

As discussed, the strategy currently has limitations and constraints that need to be addressed for better implementation. By enhancing the data inputs and refining the testing framework, the strategy can become more robust, adaptable, and realistic. The following improvements are proposed to build on the current approach:

### A. Adaptive Risk-Aversion Parameters

The risk-aversion parameter currently used in the optimization framework is fixed, which limits the strategy's flexibility. A more dynamic approach would involve developing **adaptive risk-aversion parameters** that respond to market volatility. For instance, the strategy could adjust its risk exposure in times of market turbulence to focus on capital preservation while expanding risk capacity during stable conditions. This would make the strategy more responsive to changing market dynamics.

### B. Sophisticated Transaction Cost Models

To improve back-testing realism, creating more sophisticated transaction cost models is essential. Real-world implementation faces challenges such as trading fees, bid-ask spreads, and slippage, particularly when re-balancing a dollar-neutral portfolio. Taking these transaction costs into consideration ensures the strategy reflects realistic net returns. This enhancement will also aid in identifying trade frequency thresholds that minimize costs without compromising performance.

### C. Sector-Neutral Constraints

Introducing **sector-neutral constraints** can ensure a more balanced exposure across industries within the portfolio. By maintaining neutrality within the long-short framework, the strategy can reduce concentration risk and prevent overexposure to specific sectors. Sector neutrality also aligns the strategy with broader market conditions, allowing performance to rely solely on factor-based signals.

### D. Incorporation of Liquidity Risk Measures

Liquidity risk is particularly relevant for mid-cap stocks, which often experience lower trading volumes. Integrating liquidity risk measures into portfolio construction can improve real-world implementation. This can involve filtering stocks based on minimum liquidity thresholds or optimizing trade sizes to avoid excessive market impact. Addressing liquidity concerns will enhance the portfolio's stability and scalability.

## VIII. Conclusion

In this study, the high Sharpe ratios were achieved through a market-neutral strategy highlighting the potential of mid-cap investing. We developed a robust long-short equity strategy tailored for mid-cap stocks, delivering consistent risk-adjusted returns. By leveraging data spanning from 2013 to 2023, sourced from WRDS and adhering to point-in-time (PIT) compliance, the methodology ensured transparency and reproducibility. Key financial metrics such as profitability, valuation, and liquidity were engineered to drive portfolio optimization. Backtesting across diverse market conditions demonstrated the stability and robustness of the strategy. Although our strategy returned promising results for mid-cap investing, we have also discussed potential shortcomings of our model (some more far-fetched than others) like sensitivity towards the risk-aversion parameter, lack of back-testing on Extreme Market Scenarios and more. In summary, this research emphasizes the potential of mid-cap stocks as an attractive investment avenue, blending transparency with high performance in portfolio management.

## Appendix

**Implementation Details:** The implementation is divided into two parts. The first part focuses on data extraction from WRDS and Compustat, ensuring the dataset is comprehensive and point-in-time (PIT) compliant. The second part involves the analysis itself, where key financial features are derived and systematically evaluated to optimize the portfolio strategy. Full implementation details, including code, are available on GitHub: GitHub Link to Implementation

**Final Dataset Feature Description:** The final dataset had 41 columns, containing data of stocks ranging from the year 2013 to 2023. The description of each of the columns has been provided on the following page in a tabular format for reference.

## References

[1] H. Gurani, "Predicting stock performance in Indian Mid-Cap and Small-Cap firms: An exploration of financial ratios through logistic regression analysis," *Contabilitatea, Expertiza Și Auditul Afacerilor*, vol. 4, no. 9, pp. 56–63, 2023. doi: 10.37945/cbr.2023.09.06.

[2] K. Madhukumar and A. Ravichandran, "A comparative performance evaluation between large-cap and mid-cap mutual fund returns," *Developing Country Studies*, vol. 8, no. 8, pp. 111–116, 2018. ISSN: 2224-607X (Paper), ISSN: 2225-0565 (Online).

[3] J. G. Nicholas, *Market-Neutral Investing: Long/Short Hedge Fund Strategies*. Bloomberg Press, 2000, pp. 383–417.

[4] American Century Investments, "The case for mid-cap investing: A comprehensive analysis of performance trends," 2024. [Online]. Available: https://www.americancentury.com

[5] AllianceBernstein, "Market-neutral strategies: A guide to long/short investing," 2012. [Online]. Available: https://www.alliancebernstein.com

[6] J. Johannessen and M. Nordestedt, "The effectiveness of market-neutral strategies in equity investing," *Journal of Financial Research*, vol. 46, no. 2, pp. 245–268, 2023.

[7] Kayne Anderson Rudnick, "Mid-cap investing: Opportunities and performance metrics," 2023. [Online]. Available: https://www.kayne.com

[8] P. Lynch, "The mid-cap anomaly: A comparative analysis of stock performance," *Journal of Portfolio Management*, vol. 44, no. 1, pp. 12–23, 2018.

[9] V. DeMiguel, L. Garlappi, and R. Uppal, "Optimal versus naive diversification: How inefficient is the 1/N portfolio strategy?," *The Review of Financial Studies*, vol. 22, no. 5, pp. 1915–1953, 2009.

[10] S. Basu and G. Waymire, "Costs of equity and earnings attributes," *The Accounting Review*, vol. 79, no. 4, pp. 967–993, 2004. doi: 10.2308/accr.2004.79.4.967.

[11] Panagora Asset Management, "Aspect-constrained portfolios: A framework for risk management and investment strategy development," n.d. [Online]. Available: https://www.panagora.com/assets/aspect_constrain_portfolios.pdf

[12] H. Gurani and S. Gupta, "An empirical study on the performance of mid-cap stocks in India: Evidence from the Indian stock market," *Academia Journal of Management Science*, vol. 12, no. 7, pp. 31–40, 2023. [Online]. Available: https://www.indianjournals.com

[13] R. Michaely and K. L. Womack, "The effects of analyst coverage on stock returns: Evidence from the SSRN database," *SSRN Electronic Journal*, 2010. doi: 10.2139/ssrn.315619.

TABLE II: Final Dataset Feature Description

| Data Field | Description |
| --- | --- |
| permno | Permanent number assigned to a security by CRSP, uniquely identifying the security. |
| date | The specific date of the observation. |
| prc | Closing price of the security on the date. |
| shrout | Number of outstanding shares for the company. |
| market_cap | Market capitalization, calculated as prc × shrout. |
| ret | Total return of the security, including dividends, over the period. |
| retx | Return of the security excluding dividends. |
| gvkey | Global company key, a unique identifier assigned by Compustat. |
| datadate | Date associated with the financial data, typically the fiscal year-end or quarter-end. |
| tic | Ticker symbol of the company. |
| at | Total assets of the company. |
| lt | Total liabilities of the company. |
| ceq | Common equity or shareholders' equity of the company. |
| revt | Total revenue or sales of the company. |
| gp | Gross profit, calculated as revenue minus cost of goods sold (COGS). |
| oiadp | Operating income after depreciation. |
| ni | Net income of the company. |
| act | Current assets of the company. |
| lct | Current liabilities of the company. |
| dltt | Long-term debt of the company. |
| dlc | Current portion of long-term debt or short-term debt. |
| che | Cash and cash equivalents of the company. |
| xint | Interest expense of the company. |
| ebitda | Earnings before interest, taxes, depreciation, and amortization. |
| epspx | Earnings per share (basic) excluding extraordinary items. |
| pe_ratio | Price-to-earnings ratio. Indicates valuation based on earnings, useful for growth vs. value analysis. |
| pb_ratio | Price-to-book ratio. Reflects valuation relative to book value, critical for value-based strategies. |
| ps_ratio | Price-to-sales ratio. Measures valuation against sales, useful for high-growth companies. |
| enterprise_value | Total valuation of the company, calculated as market cap + debt - cash. Represents the firm's true value. |
| ev_to_ebitda | Enterprise value / EBITDA, a measure of valuation relative to earnings. |
| gross_margin | Gross profit as a percentage of revenue. Reflects profitability before other expenses are deducted. |
| operating_margin | Operating income as a percentage of revenue. Indicates operational efficiency. |
| net_margin | Net income as a percentage of revenue. Indicates overall profitability after all expenses. |
| current_ratio | Ratio of current assets to current liabilities. Indicates ability to meet short-term obligations. |
| debt_to_equity | Ratio of total debt to shareholders' equity. |
| interest_coverage | Ratio of EBITDA to interest expense, a measure of solvency. |
| linktype_y | Type of link between datasets (e.g., security-to-firm link). |
| linkprim_y | Primary link indicator, showing whether the link is primary. |
| linkdt_y | Start date of the link between datasets. |
| linkenddt_y | End date of the link between datasets. |
| avg_sentiment | Average sentiment score from textual analysis of news, filings, or other documents. |